\documentclass[twocolumn,prl,superscriptaddress,showpacs,aps,10pt,footinbib,longbibliography]{revtex4-2}
\newcommand{\pagenumbaa}{1}
\usepackage{amssymb,amsmath}   
\usepackage[dvips]{graphicx}   
\usepackage{verbatim}   
\usepackage{color}      
\usepackage{subfigure}  
\usepackage{hyperref}   
\usepackage{gensymb}
\usepackage{epstopdf}
\usepackage{natbib}
\usepackage{enumerate}
\usepackage{ulem}
\usepackage{braket}

\DeclareUnicodeCharacter{2061}{}

\newcommand{\Tone}{{T_\mathrm{1}}}
\newcommand{\be}{\begin{equation}}
\newcommand{\ee}{\end{equation}}

\newif\ifspectro
\spectrofalse

\usepackage{units}
\begin{document}


\title{Experimentally revealing anomalously large dipoles in a quantum-circuit dielectric}


\author{Liuqi Yu}
\thanks{Correspondence and requests for materials should be addressed to L.Y. (email: liuqi.yu.physics@gmail.com) and K.D.O (email: osborn@lps.umd.edu).}
\affiliation{Laboratory for Physical Sciences, University of Maryland, College Park, MD 20740, USA}
\affiliation{Department of Physics, University of Maryland, College Park, MD 20742, USA}

\author{Shlomi Matityahu}
\affiliation{Department of Physics, Ben-Gurion University of the Negev, Beer Sheva 84105, Israel}
\affiliation{Institut f\"ur Theorie der Kondensierten Materie, Karlsruhe Institute of Technology, 76131 Karlsruhe, Germany}

\author{Yaniv J. Rosen}
\affiliation{Lawrence Livermore National Laboratory, Livermore, California 94550 USA}

\author{Chih-Chiao Hung}
\affiliation{Laboratory for Physical Sciences, University of Maryland, College Park, MD 20740, USA}
\affiliation{Department of Physics, University of Maryland, College Park, MD 20742, USA}

\author{Andrii Maksymov}
\affiliation{Department of Chemistry, Tulane University, New Orleans, LA 70118, USA}

\author{Alexander L. Burin}
\affiliation{Department of Chemistry, Tulane University, New Orleans, LA 70118, USA}

\author{Moshe Schechter}
\affiliation{Department of Physics, Ben-Gurion University of the Negev, Beer Sheva 84105, Israel}

\author{Kevin D. Osborn}
\affiliation{Laboratory for Physical Sciences, University of Maryland, College Park, MD 20740, USA}
\affiliation{Joint Quantum Institute, University of Maryland, College Park, MD 20742, USA}


\maketitle

\setcounter{page}{\pagenumbaa}
\thispagestyle{plain}


\twocolumngrid
\textbf{
Quantum two-level systems (TLSs) intrinsic to glasses induce decoherence in many modern quantum devices, such as superconducting qubits. Although the low-temperature physics of these TLSs is usually well-explained by a phenomenological standard tunneling model of independent TLSs, the nature of these TLSs, as well as their behavior out of equilibrium and at high energies above $1\,$K, remain inconclusive. Here we measure the non-equilibrium dielectric loss of TLSs in amorphous silicon using a superconducting resonator, where energies of TLSs are varied in time using a swept electric field. Our results show the existence of two distinct ensembles of TLSs, interacting weakly and strongly with phonons, where the latter also possesses anomalously large electric dipole moment. These results may shed new light on the low temperature characteristics of amorphous solids, and hold implications to experiments and applications in quantum devices using time-varying electric fields.}

\begin{figure*}
 \center
  \includegraphics[width=\textwidth]{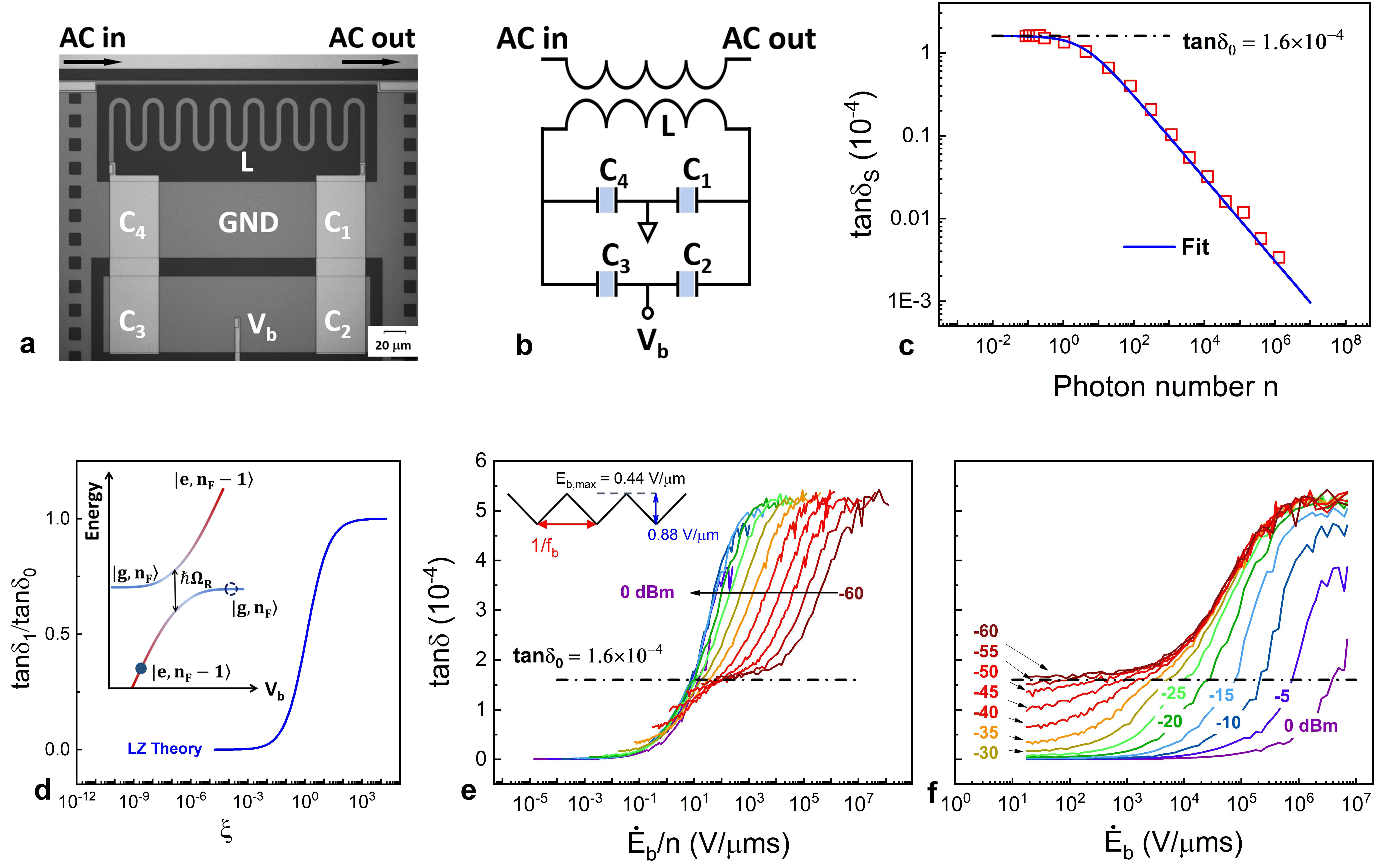}\vspace{-3mm}
\caption{
\textbf{Steady-state and nonequilibrium loss.} \textbf{a}~Optical image and \textbf{b}~schematic of the superconducting resonator device, where the total capacitance, $C$, is made of a bridge of 4 equal capacitances ($C_1=C_2=C_3=C_4 = C$). Bias voltage $V_{\rm b}$ is applied across $C_1$ and $C_2$ (and also $C_3$ and $C_4$) in series. The inductor $L$ is coupled to a coplanar waveguide for transmission measurements. \textbf{c}~Steady-state loss tangent $\tan\delta_{\rm s}$ (red squares) plotted as a function of the average photon number $n$ at 20\,mK. The loss fits to the standard model of TLS loss: $\tan\delta_{\rm s}=\tan\delta_0/\sqrt{1+n/n_{\rm c}}$ (blue), where $\tan\delta_{0}=1.6\times10^{-4}$ is the intrinsic material loss measured at the single photon regime ($n<1$) and $n_{\rm c} = 3.7$, is the quantum-classical crossover photon number. \textbf{d}~Calculation of normalized non-equilibrium loss $\tan\delta/\tan\delta_0$ as a function of the dimensionless bias rate  $\xi=2v_0/(\pi\Omega^{2}_{\rm{R}0})=\frac{2\varepsilon\mathbb{V}}{\pi\omega_{0}p}\cdot\frac{\dot{E}_{\rm b}}{n}$, based on the STM in the regime of strong saturation ($\Gamma_{1,\rm m}/\Omega_{\rm{R}0}\ll 1$)~\cite{Burin2013}. Inset: TLS-photon energy-level diagram as a function of the TLS energy bias rate $v$. The states $|g,n_{\rm F}\rangle$ and $|e,n_{\rm F}-1\rangle$ ($|g\rangle$ and $|e\rangle$ are the TLS ground and excited states, and $|n_{\rm F}\rangle$ is a photon number state) are connected by an adiabatic transition (with an avoided crossing gap equal to the TLS Rabi frequency $\Omega_{\rm{R}}$) which may lead to single-photon loss; the probability for photon absorption is $1-e^{-\pi\Omega^{2}_{\rm{R}}/(2v)}$. \textbf{e}~Loss measured at various microwave source powers (-60\,dBm to 0\,dBm with 5\,dBm increment as indicated by the arrow) as a function of $\dot{E}_{\rm b}/n$ (proportional to $\xi$). A periodic triangular bias voltage is applied, as indicated by the inset, with a fixed amplitude $E_{\rm b,max}=0.44\,$V$/\mu$m and varying modulation frequency $f_{\rm b}$, resulting in the bias rate $\dot{E}_{\rm b}=4E_{\rm b,max}f_{\rm b}$. Data collapse at medium bias rates occurs below the intrinsic loss $\tan\delta_{0}$ (dashed-dotted black line), which scales according to LZ theory. The scaling deviates at the lowest rates, where TLS relaxation ($\Tone$) processes dominate. Excess loss above $\tan\delta_{0}$ can be seen at $\dot{E}_{\rm b}/n\gtrsim 10\,$V$/\mu$m$\,$s. \textbf{f}~ Electric field bias rate dependence of $\tan\delta$ at selected microwave source powers. Excess loss above $\tan\delta_{0}$ can be seen at fast bias rates, which reveals a data collapse at low powers $P_{\rm ac}<-25\,$dBm. \vspace{-4mm}}
\label{Figure:1}
\end{figure*}

\noindent
Two-level system (TLS) defects are known to decohere qubits\cite{Martinis2005, Wang2015}. Furthermore, there is an effort to control their relaxation\cite{Klimov2018}, and understand their dephasing\cite{Burnett2019} for future quantum information processors. TLS are also investigated as the focus of single-photon studies which extract their information, including their decoherence and couplings to other quantum elements\cite{Muller2019,Oliver2013,Martinis2005,Shalibo2010,Zaretskey2013,Christensen2019,Pappas2011,Paik2010,Muller2015,Klimov2018,Schlor2019,Burnett2019,deGraaf2018,Gao2008,Burnett2014,Faoro2015,Burin2015,Connors2019}. The science of TLSs has uncovered important phenomena including how noise is created on TLSs that are close to the frequency of qubits\cite{deGraaf2018}. However, there is also the possibility that there are new opportunities\cite{Rosen2016,Matityahu2019} or challenges ahead posed by TLSs because it is still unresolved how much qubit relaxation may result out of equilibrium. Specifically, the active use of a voltage gate in superconducting\cite{Casparis2018}, semiconducting\cite{Kawakami2014}, and Majorana\cite{Plugge2017} qubits could cause unexpected changes in qubit coherence because time variance can reveal unexpected scientific results.

The properties of dielectrics in qubits, and amorphous materials at low energies $T\lesssim 1\,$K\cite{Zeller1971}, are generally attributed to atoms or groups of atoms that can tunnel between two adjacent structural configurations. They can be described by a standard tunneling model (STM)\cite{Phillips1972,Anderson1972}, which assumes the existence of independent TLSs with asymmetry energy $\Delta$ and tunneling amplitude $\Delta_0$ that are distributed universally, e.g., approximately describing all amorphous dielectrics, according to a distribution function $P(\Delta,\Delta_0)\propto P_0/\Delta_0$, where $P_0$ is a material constant. This leads to an energy independent density of states (DOS) as a function of the energy splitting $E=\sqrt{\Delta^{2}+\Delta^{2}_{0}}$~\cite{Phillips1972,Anderson1972,Phillips1987,Hunklinger1986}. This model accounts well for most of the low energy properties observed in a broad range of amorphous systems\cite{Phillips1987,Hunklinger1986,Pohl2002}. Within the STM the strength of the TLS-phonon coupling $\gamma$ relates to the dimensionless tunneling strength, $C_0=P_0 \gamma^2/(\rho v^2)$, where $\rho$ is the mass density and $v$ the acoustic velocity. While the STM does not specify the value of the tunneling strength, experimentally it is found to be universally small, of order $10^{-3}$, leading to weak and remarkably universal phonon attenuation properties in amorphous solids. Recent studies on material density\cite{Queen2015,Molina2020,Liu2021}, stress\cite{Southworth2009} and TLS-nucleus interactions\cite{Bartkowiak2013} extend our knowledge of TLS origins. However, an understanding of the nature of TLSs and of the origin of low-temperature universality in the phonon attenuation of glasses is still lacking~\cite{Pohl2002,Yu1988,Liu1997}. 

Individual TLSs have been studied via adiabatic adjustments of electric fields, and adjustments of strain field. The external field changes the TLS asymmetry energy $\Delta$, which in turn modifies the TLS energy splitting $E=\sqrt{\Delta^{2}+\Delta^{2}_{0}}$, while the tunneling energy $\Delta_0$ is fixed. Using static strain and electric fields to tune TLSs in qubits or resonators provides a spectroscopy method to extract energy splittings~\cite{Grabovskij2012}, couplings to phonons~\cite{Grabovskij2012,Brehm2017}, dipole moments~\cite{Sarabi2016,Lisenfeld2019,Hung2022}, relaxation and decoherence rates~\cite{Lisenfeld2016} of individual TLSs, and recently even their locations~\cite{Bilmes2020,Bilmes2021}. These experiments probe TLSs in resonance with the device in equilibrium conditions at $\sim 5\,$GHz, corresponding to energies of $\sim k_{\rm B}$(0.25K), where $k_{\rm B}$ is the Boltzmann constant. Consequently, good agreement with the STM predictions is found in the majority of the cases. 

In a related study to this work\cite{Khalil2014}, TLSs were swept through resonance with a resonator and emit the energy afterward as a phonon. By using fast-bias rates and large bias amplitudes, this technique allows the study of TLSs out of equilibrium, probing TLSs which are at high energies, but are temporarily lowered to the single-photon energy at the resonator. For a past measurement of amorphous silicon nitride\cite{Khalil2014}, the bias rate dependence of the non-equilibrium dielectric loss is explained well by the STM and the Landau-Zener (LZ) effect\cite{Burin2013} -- it describes the scaling of non-equilibrium loss over a range of ac-field amplitudes and (swept) bias rates. The TLS dipole moment is then extracted from the data during scaling. It is well known that at equilibrium (zero bias field), the TLS microwave absorption is saturated at high driving powers (average photon number $n\gg 1$) and the resonator loss decreases as a function of the driving power, as already shown five decades ago~\cite{VonSchickfus1977}. More recently, it was shown that a time-dependent bias electric field $E_{\rm b}(t)$ sweeps TLSs into resonance, with a rate proportional to $\dot{E}_{\rm b}$\cite{Burin2013,Khalil2014}. The dielectric loss increases with $\dot{E}_{\rm b}$ up to the intrinsic loss tangent $\tan\delta_0$, i.e. the loss in the equilibrium ($\dot{E}_{\rm b}=0$) and low-power limit $n\lesssim 1$, where TLSs are unsaturated. It should be noted that the observations were made previously in silicon nitride\cite{Khalil2014}, but are expected to apply to all amorphous materials where the STM holds. A similar fast-bias technique has been used to demonstrate a defect maser based on TLS population inversion~\cite{Rosen2016} and dynamical decoupling of TLSs from a resonator by multiple coherent resonant transitions~\cite{Matityahu2019}. 

An application of the STM to material data generally leads to the conclusion of small interactions between TLSs, related to an energy independent density of states. Early work on non-equilibrium measurements of dielectrics revealed small amplitude relaxations~\cite{Natelson1998} which were understood from a small gap in the standard TLS density of states, formed by rare strong interactions~\cite{Burin1995}. Presently, modern measurements provide an opportunity for new discovery from the change of TLS energies at high rates. This allows fast lowering high-energy TLSs and they may undergo single-photon exchanges with a microwave resonator mode. This gives the opportunity to uncover phenomena on TLS density which is relevant to modern electric-biased quantum devices.

Here we employ the fast-bias technique of Ref.~\cite{Khalil2014}, but a different material, amorphous silicon is studied in this work. At low bias rates the dielectric loss follows the theory of Ref.~\cite{Burin2013} based on the STM, similar to previous measurements on silicon nitride~\cite{Khalil2014}. However, at high bias rates we observe a striking excess loss, a loss much larger than the intrinsic loss, for all applied driving powers in the amorphous film under study. This result contradicts the common understanding of the STM and provides strong evidence for another loss mechanism from high-energy TLSs probed out of equilibrium. We analyze our data in terms of two types of TLSs, as proposed in Ref.~\cite{Schechter2013}, where the first type interacts weakly while the second interacts strongly with phonons, with their own contributions to the total loss. To our knowledge, our observations provide the first direct experimental evidence for the existence in amorphous solids of two types of TLSs with coupling to phonons that differ by an order of magnitude.
\\

\noindent\textbf{Results}\\
\noindent\textbf{Nonequilibrium excess loss.} As shown in Fig.~\ref{Figure:1}a and Fig.~\ref{Figure:1}b, the resonator consists of four equal bridge parallel-plate capacitors with a total capacitance, $C$ ($C_1=C_2=C_3=C_4=C$). A dc bias field, $E_{\rm b}$, is applied across the capacitors. The bridge layout is effective to isolate the ac resonance energy from the bias field input port. Standard transmission measurements of $S_{21}$ are performed on the resonator as a function of the average photon number, $n$, or microwave resonator field amplitude $E_{\rm ac}=\sqrt{2n\hbar\omega_0/(\varepsilon\mathbb{V})}$, where $\hbar$ is the reduced Planck constant, $\omega_0=2\pi\times5.1\,$GHz is the resonator resonance frequency, $\varepsilon$ is the permittivity, and $\mathbb{V}=2925\,\mu$m$^{3}$ is the total capacitor volume. The material loss tangent $\tan\delta$, equal to the inverse resonator quality $1/Q_{\rm i}$, is then extracted. Figure~\ref{Figure:1}c shows the steady state loss tangent, $\tan\delta_{\rm s}$, at zero bias field as a function of $n$. The resonator photon power dependence arises from the saturation of TLSs, and the loss tangent $\tan\delta_{\rm s}$ is expressed as $\tan\delta_{\rm s}=\tan\delta_0/\sqrt{1+n/n_{\rm c}}$~\cite{VonSchickfus1977}, where $\tan\delta_0$ is the intrinsic material loss measured in the single photon limit ($n \ll 1$) and $n_{\rm c}$ is the quantum-classical crossover photon number. The fit yields $n_{\rm c}=3.7$ and $\tan\delta_0=1.6\times 10^{-4}$, corresponding to $Q_{\rm i} = 6200$.

When the bias field $E_{\rm b}$ is varied in time (see inset of Fig.~\ref{Figure:1}e), the asymmetry energy of each TLS is modified as $\Delta(t)=\Delta(0)-2pE_{\rm b}(t)\cos\theta$, where $p$ is the TLS dipole moment and $\theta$ is the angle between $\vec{p}$ and $\vec{E}_{\rm b}$. An ensemble of ground-state TLSs is swept through the resonance, described by the condition $E(t)\equiv\sqrt{\Delta^{2}_{0}+\Delta^{2}(t)}=\hbar\omega_0$ for each TLS, thereby leading to an enhanced loss. Close to resonance, the TLS energy changes at a rate of $v=|\dot{E}|/\hbar\approx v_0\sqrt{1-(\Delta_0/\hbar\omega_0)^{2}}\cos\theta$, where $v_0=(2p/\hbar)\dot{E}_{\rm b}$ is the maximum bias rate. The dynamics of each resonant passage is of the LZ type, where an adiabatic transition corresponds to the excitation of a TLS by absorption of a single photon (see inset of Fig.~\ref{Figure:1}d). The resulting non-equilibrium loss was analyzed in Ref.~\cite{Burin2013} and is shown to be a function of the two dimensionless parameters: the standard LZ parameter $\xi\equiv 2v_0/(\pi\Omega^{2}_{\rm{R}0})$ and the ratio $\Gamma_{1,\rm m}/\Omega_{\rm{R}0}$, where $\Omega_{\rm{R}0}=pE_{\rm ac}/\hbar$ is the maximum TLS Rabi frequency and $\Gamma_{1,\rm m}$ is the maximum TLS relaxation rate. As plotted in Fig.~\ref{Figure:1}d, in the regime of strong saturation $\Gamma_{1,\rm m}/\Omega_{\rm{R}0}\ll 1$, the predicted normalized loss $\tan\delta/\tan\delta_0$ is a universal function of $\xi$, approaching 1 in the non-adiabatic limit $\xi\gg 1$. It should be noted that the intrinsic loss is recovered at high bias rates provided that the TLS DOS is energy-independent, as assumed by the STM. This is understood as a result of fast LZ passage time compared to $2\pi/\Omega_{\rm{R}0}$, such that the swept TLSs remain unsaturated by the resonator field, as in the single-photon limit.

Fig.~\ref{Figure:1}e shows the measured dielectric loss tangent $\tan\delta$ as a function of $\dot{E}_{\rm b}/n$, which is proportional to the dimensionless bias rate $\xi=\frac{2\varepsilon\mathbb{V}}{\pi\omega_{0}p}\cdot(\dot{E}_{\rm b}/n)$. The data demonstrate two distinct loss regimes separated by $\tan\delta_0$. Below $\tan\delta_0$ the curves show a single dependence on $\dot{E}_{\rm b}/n$ for all microwave driving powers, except for small deviations at small $\dot{E}_{\rm b}/n$ (or $\xi\lesssim\Gamma_{1,\rm m}/\Omega_{\rm{R}0}$) due to incoherent LZ transitions~\cite{Khalil2014,Burin2013}. The data collapse of the different curves is in accord with the theory of Ref.~\cite{Burin2013} and agrees with previous measurements in silicon nitride~\cite{Khalil2014}, suggesting that dielectric loss originates from standard TLSs. However, at higher values of $\dot{E}_{\rm b}/n$, $\tan\delta$ strikingly exceeds the intrinsic loss $\tan\delta_0 = 1.6\times 10^{-4}$ and reaches a maximum of $5.2\times10^{-4}$ except for the high powers, $P_{\rm ac}>-20\,$dBm (see Fig.~\ref{Figure:1}f). The excess loss scales with $\dot{E}_{\rm b}/n$ at high driving powers, $P_{\rm ac}>-20\,$dBm (see Fig.~\ref{Figure:1}e), but scales with $\dot{E}_{\rm b}$ at low driving powers, $P_{\rm ac} < -30\,$dBm (see Fig.~\ref{Figure:1}f). These distinguished saturation behaviors at small and large driving powers imply that the excess loss is due to a second type of TLSs. This saturation occurs at much higher driving powers ($P_{\rm ac}\sim -30\,$dBm) compared to $\sim -60\,$dBm for the saturation of standard TLSs responsible for the loss at low bias rates. It therefore suggests that the second type of TLSs has much higher relaxation rates. In addition, at small powers $P_{\rm ac}<-25\,$dBm where the second type of TLSs are unsaturated, the loss shows a bias rate dependence (Fig.~\ref{Figure:1}f), indicating a rate-(thus energy-) dependent DOS of the second type of TLSs. In comparison, the non-equilibrium loss of unsaturated standard TLSs is equal to the intrinsic value $\tan\delta_0$, irrespective of the bias rate.
\\

\begin{figure}[!htb]
\includegraphics[width=1\columnwidth]{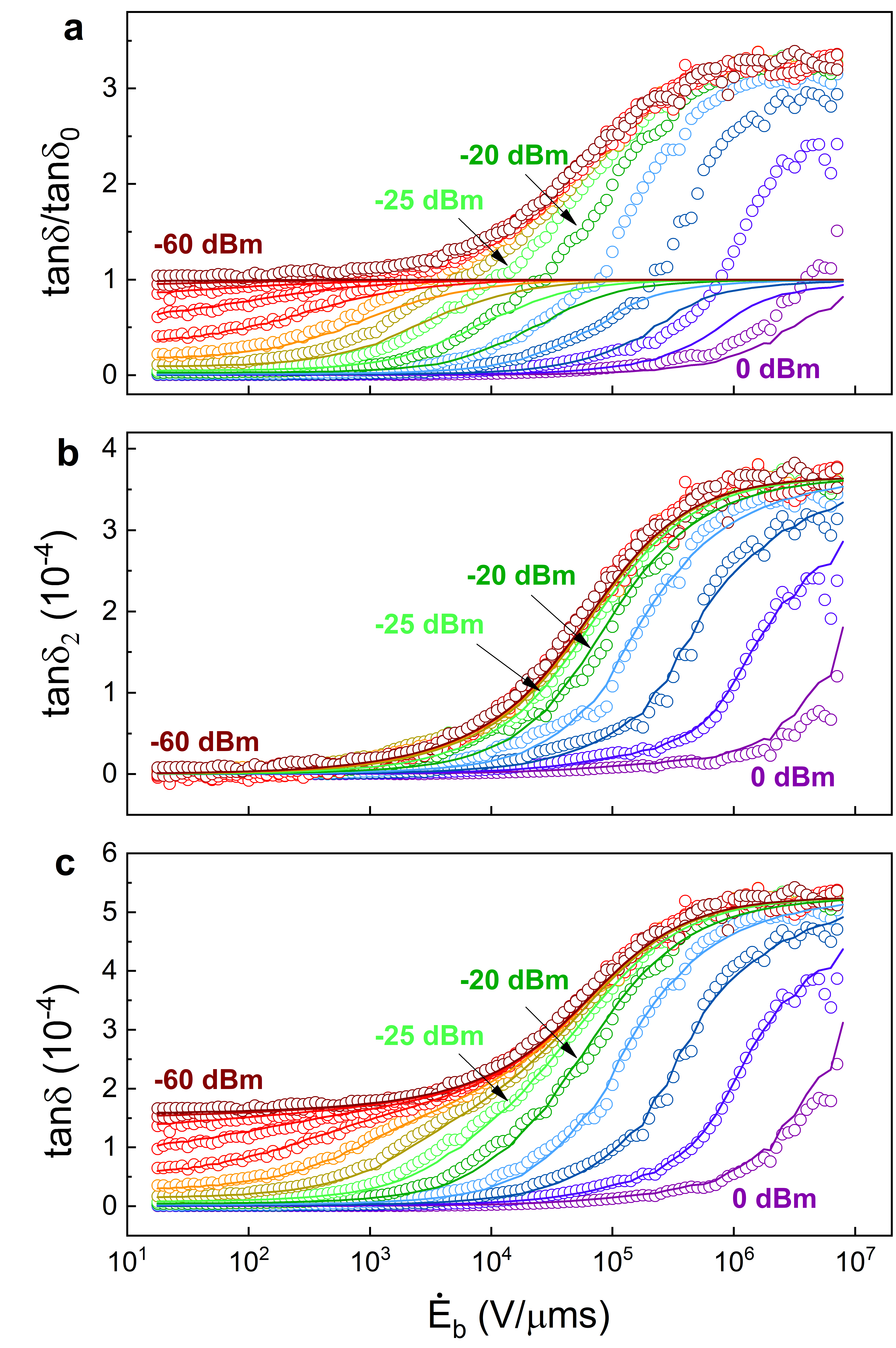}\vspace{-3mm}
\caption{
\textbf{Loss analysis of the first and the second types of TLSs.} \textbf{a}~Normalized loss $\tan\delta/\tan\delta_0$ (colored circles), where $\tan\delta_0=1.6 \times 10^{-4}$ is the intrinsic loss at low powers and zero bias rate, plotted at each driving power (in dBm). Colored solid curves are calculations based on LZ theory within the STM~\cite{Burin2013}, performed at low bias rates $\dot{E}_{\rm b}<300\,$V$/\mu$m$\,$s. The calculation gives $p_1=11\,$D ($1\,D = 0.21\,e\text{\r{A}}$, where $e$ is the electron charge) and a maximum relaxation rate $\Gamma^{(1)}_{1,\rm m}=5.7\,$MHz for the standard (first) TLS type. \textbf{b}~The excess loss $\tan\delta_2=\tan\delta-\tan\delta_1$ (colored circles) is attributed to the second type of TLSs. Colored solid curves are theoretical calculations of $\tan\delta_2$ based on an energy-dependent DOS for the second TLS type. \textbf{c}~Loss data (colored circles) corresponding to Fig.~\ref{Figure:1}f are shown together with the theoretical calculations (colored solid lines), which combine both types of TLS loss. \vspace{-4mm}}
\label{Figure:2}
\end{figure}

\noindent\textbf{Loss from two types of TLSs.} We now analyze the data in a model which consists of two types of TLSs. First, we use the data at low to intermediate bias rates to calculate the loss resulting from the standard TLSs, $\tan\delta_1$. Figure~\ref{Figure:2}a shows the loss of Fig.~\ref{Figure:1}f normalized by $\tan\delta_0$, along with a numerical calculation of $\tan\delta/\tan\delta_0$ based on LZ theory (see Methods and Refs.~\cite{Khalil2014,Burin2013} for details). The calculated loss at a given bias rate $\dot{E}_{\rm b}$ and photon number $n$ depends on the dipole moment $p$ and maximum relaxation rate $\Gamma_{1,\rm m}$ which serve as fitting parameters. We conduct the fit in the low bias rate regime $\dot{E}_{\rm b}<300\,$V$/\mu$m$\,$s and obtain $p_1=11\,$D and $\Gamma^{(1)}_{1,\rm m}=5.7\,$MHz for the standard TLSs. In comparison to quasi-static tuned TLS measurements of amorphous silicon nitride\cite{Sarabi2016} and thin-film crystalline alumina\cite{Hung2022}, the total average dipole moment $p_z$ is 3-4 D, leading to a total representative moment of approximately $p = 6-7\,$D.  In two cases, the extracted moment seems larger. In dynamical tuned measurements of silicon nitride\cite{Khalil2014}, a single representative dipole moment was extracted as $p = 7.9\,$D, and this may partially be larger due to the different technique, rather than the material difference. Additionally, in amorphous alumina, some very large moments of $p_z \sim 10\,$D are observed\cite{Hung2022} and a delocalized oxygen model might account for these unconventionally large moments\cite{DuBois2013}. We thus find that the first TLS moment type extracted from our amorphous silicon data, $p_1\approx 11\,$D, to be within an expected range of standard TLSs, given large range of analyzed moments.

The calculations generally capture the loss well but show increasing variance with increasing $\dot{E}_{\rm b}$, since the contribution of the second type of TLSs becomes increasingly significant at higher bias rates. In Fig.~\ref{Figure:2}b we plot the net excess loss $\tan\delta_2$ by subtracting the calculated standard TLS loss from the measured loss, i.e.\ $\tan\delta_2=\tan\delta-\tan\delta_1$. The single dependence on $\dot{E}_{\rm b}$ for low driving powers $P_{\rm ac}<-25\,$dBm then becomes apparent. The independence on driving power implies that the TLSs responsible for the excess loss are unsaturated in this power regime, such that the observed single curve is the equivalent of the intrinsic loss $\tan\delta_0$ of the standard TLSs, and will be denoted as $\tan\delta_{2,0}$. The fact that $\tan\delta_{2,0}$ increases with $\dot{E}_{\rm b}$ points to two features of the DOS of the contributing TLSs. First, the DOS is an increasing function of the TLS energy, because the loss is determined by the number of TLSs within the energy range $\hbar\omega_0<E<\hbar\omega_0+pE_{\rm b,max}$ that are swept through resonance. Second, Since $\dot{E}_{\rm b}$ is varied by varying the modulation frequency $f_{\rm b}$ with a \textit{fixed} amplitude $E_{\rm b,max}$, the initial energies of the TLSs that are swept through resonance are independent of $\dot{E}_{\rm b}$. This means that the large non-equilibrium DOS of these TLSs at the resonance energy $\hbar\omega_0$ tends to restore its small equilibrium value by some mechanism. This mechanism becomes less effective as the sweep time reduces ($\dot{E}_{\rm b}$ increases).

Such a scenario arises in a previously proposed two-TLS model, which divides TLSs into two groups, distinguished by their interactions with phonons~\cite{Schechter2013}. As a consequence of their distinct interactions, TLSs that are weakly coupled to phonons are abundant at low energies below $\sim1\,$K and form the standard TLSs of the STM with an approximately energy-independent DOS $\rho_1$; TLSs that are strongly coupled to phonons are characterized by an energy-dependent DOS $\rho_2(E)$ exhibiting a soft (power-law) gap at low energies. This soft gap is a result of their mutual interactions with the standard TLSs~\cite{Churkin2014,Churkin2014_2}, as dictated by the Efros-Shklovskii mechanism for long-range interacting particles in glassy systems~\cite{Efros1975,Baranovski79}. As a result, strongly interacting (with phonons) TLSs are scarce at low energies, thus rarely observable in conventional measurements performed near equilibrium. However, in our measurement TLSs with maximum energy of $\hbar\omega_0+pE_{\rm b,max}$, where their densities are much larger, can be swept into resonance. Out of equilibrium, the interaction between the two types of TLSs acts to reform the equilibrium gap in the DOS of strongly interacting TLSs by rearrangement of the standard TLSs. This sets a typical time scale for the reformation of the gap, equal to a typical relaxation time of standard TLSs. One observes the excess loss when the bias field modulation frequency $f_{\rm b}$ exceeds the standard TLS relaxation rate, such that the reconstruction of the gap is incomplete. The non-equilibrium DOS $\rho_2(E,\dot{E}_{\rm b})$ depends on energy due to the energy dependence of the equilibrium DOS $\rho^{}_{2,\rm{eq}}(E)$, and also depends on $\dot{E}_{\rm b}$ due to the time-dependent reformation of the gap. By extending the Efros-Shklovskii argument to our non-equilibrium situation we obtain an approximated expression for $\rho_2(E,\dot{E}_{\rm b})$ (see Methods for details). The resulting non-saturated loss tangent $\tan\delta_{2,0}(\dot{E}_{\rm b})=\pi p^{2}_{2}\rho^{}_{2}(\hbar\omega_0,\dot{E}_{\rm b})/(3\varepsilon)$, where $p_2$ is the dipole moment of the second type of TLSs, reads
\begin{equation}
\label{Eq:1}
\tan\delta_{2,0}(\dot{E}_{\mathrm b})=A\,\mathrm{exp}\Bigg[B\int^{C}_{0}dx\,e^{-(D/\dot{E}_{\mathrm b})e^{2x}}\Bigg],
\end{equation}
with the four fitting parameters $A$, $B$, $C$, and $D$. Here $A=\pi p^{2}_{2}\rho^{}_{2,\mathrm{eq}}(\hbar\omega_0)/(3\varepsilon)$ is the excess loss at equilibrium ($\dot{E}_{\rm b}=0$), and is proportional to the equilibrium DOS $\rho^{}_{2,\mathrm{eq}}(\hbar\omega_0)$ at the resonance energy. $B=(8\pi/3)\rho^{}_{1}u\ln\left[1+E_0/(\hbar\omega_0)\right]$, where $E_0=\min\{E_{\rm{max}},\hbar\omega_0+p_{2}E_{\rm b,max}\}$ with $E_{\rm{max}}\sim 10\,k_{\rm B}$K being the maximum energy of standard TLSs, is proportional to the phonon-mediated interaction strength $u$ between the two types of TLSs. $C=\ln\left(E_{\rm{max}}/\Delta^{}_{0,\mathrm{min}}\right)$ is the upper limit of the integral over the normalized tunneling amplitudes of the standard TLSs $x=\ln(\Delta_0/\Delta_{0,\rm{min}})$, where $\Delta_{0,\rm{min}}$ is a minimum cutoff for tunneling amplitudes~\cite{Phillips1987,Hunklinger1986}. Finally, $D=\Gamma_{1,\rm m}(E_{\rm{max}})\left(\Delta_{0,\rm{min}}/E_{\rm{max}}\right)^{2}E_{\rm{b,max}}$, with $De^{2C}=\Gamma_{1,\rm m}(E_{\rm{max}})E_{\rm{b,max}}$ being the characteristic bias rate above which standard TLSs cannot change their state in order to equilibrate the DOS of the second type of TLSs. Fitting the collapsed data of Fig.~\ref{Figure:2}(b) (for $P_{\rm ac}<-25\,$dBm) yields $A=1.8\times 10^{-8}$, $B=1.27$, $C=7.8$ and $D=7.2\times 10^{-3}\,$V$/\mu$m$\,$s. 

\begin{figure}[!htb]
\includegraphics[width=1\columnwidth]{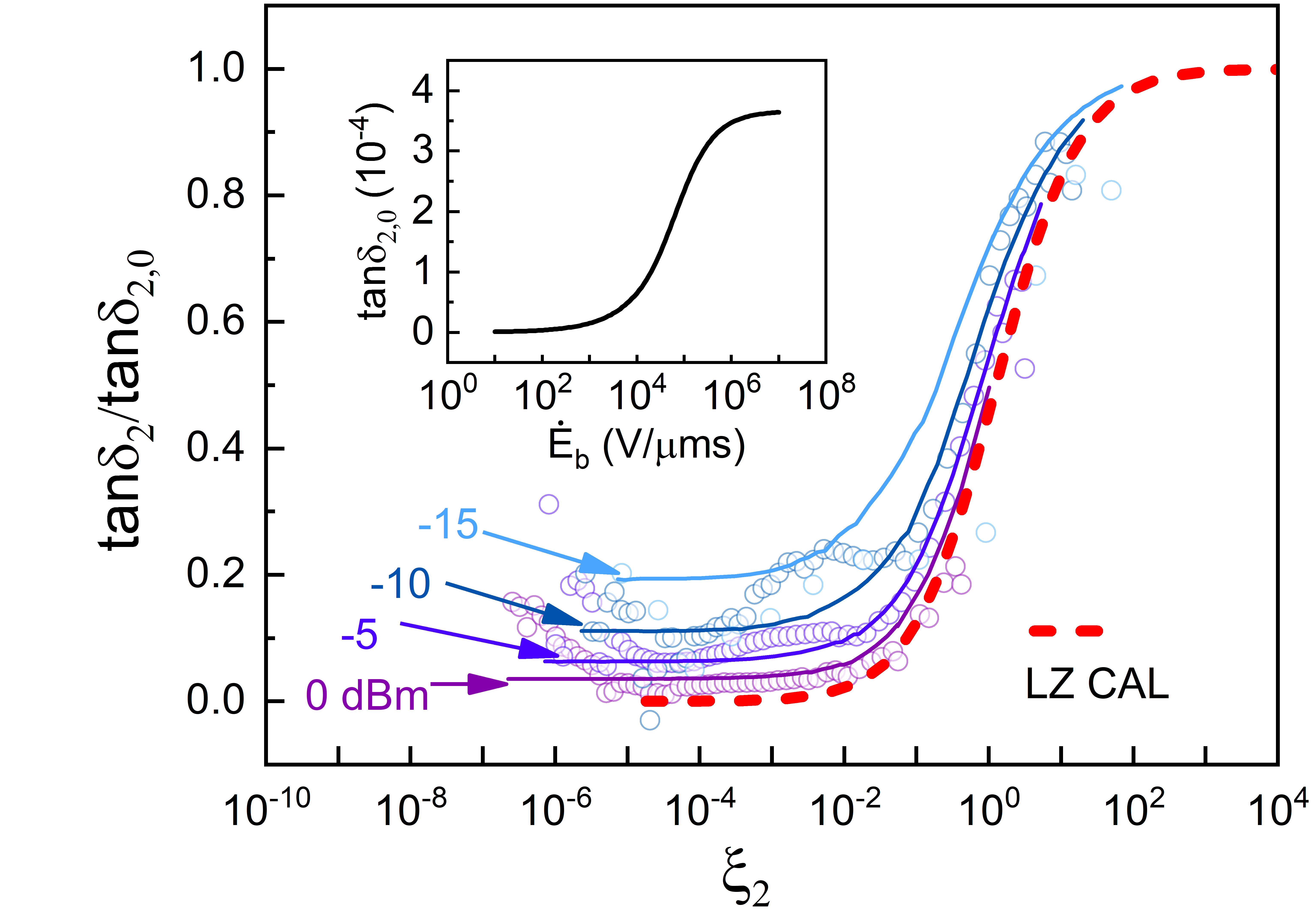}\vspace{-3mm}
\caption{
\textbf{LZ scaling for the second TLS type.} LZ analysis for the normalized excess loss $\tan\delta_2/\tan\delta_{2,0}$(colored circles), where $\tan\delta_{2,0}$ (inset) is the fit for the unsaturated excess loss at low powers $P_{\rm ac}<-20\,$dBm. The normalized excess loss at high powers ($0$ to $-15\,$dBm), where the second types of TLSs become saturated, obeys LZ scaling with the dimensionless bias rate $\xi_2=\frac{2\varepsilon\mathbb{V}}{\pi\omega_0 p_2}\cdot(\dot{E}_{\rm b}/n)$, similar to the loss due to standard TLSs [Fig.~\ref{Figure:2}a]. The numerical calculation of $\tan\delta_2/\tan\delta_{2,0}$ using LZ theory (colored solid lines) yields $p_2=110\,$D and $\Gamma^{(2)}_{1,\rm m}=800\,$MHz for the second TLS type. The LZ scaling verifies the TLS nature of the excess loss mechanism.
\vspace{-4mm}}
\label{Figure:3}
\end{figure}

At $P_{\rm ac}\geq -15\,$dBm where a saturation effect is observed, we notice that the normalized excess loss $\tan\delta_2/\tan\delta_{2,0}$, which neutralizes the effect of energy-dependent DOS of the second TLS type, resumes a simular scaling with $\dot{E}_{\rm b}/n$ as the standard TLSs (Fig.~\ref{Figure:3}). This provides a striking evidence for the TLS origin of the excess loss. We therefore repeat the numerical calculation based on LZ theory for the excess loss of Fig.~\ref{Figure:2}b, similarly to the calculation performed for the standard TLSs in Fig.~\ref{Figure:2}a. The colored solid curves shown in Fig.~\ref{Figure:2}b are obtained for $p_2=110\,$D and $\Gamma^{(2)}_{1,\rm m}=800\,$MHz for the second type of TLSs. Combining the theoretical calculations of $\tan\delta_1$ and $\tan\delta_2$ from the standard and second TLS types, respectively, the total loss $\tan\delta=\tan\delta_1+\tan\delta_2$ agrees well with measured loss over the entire domain of the driving powers and bias rates explored in the experiment, as shown in Fig.~\ref{Figure:2}c.
\\

\begin{figure}[tb]
\includegraphics[width=1\columnwidth]{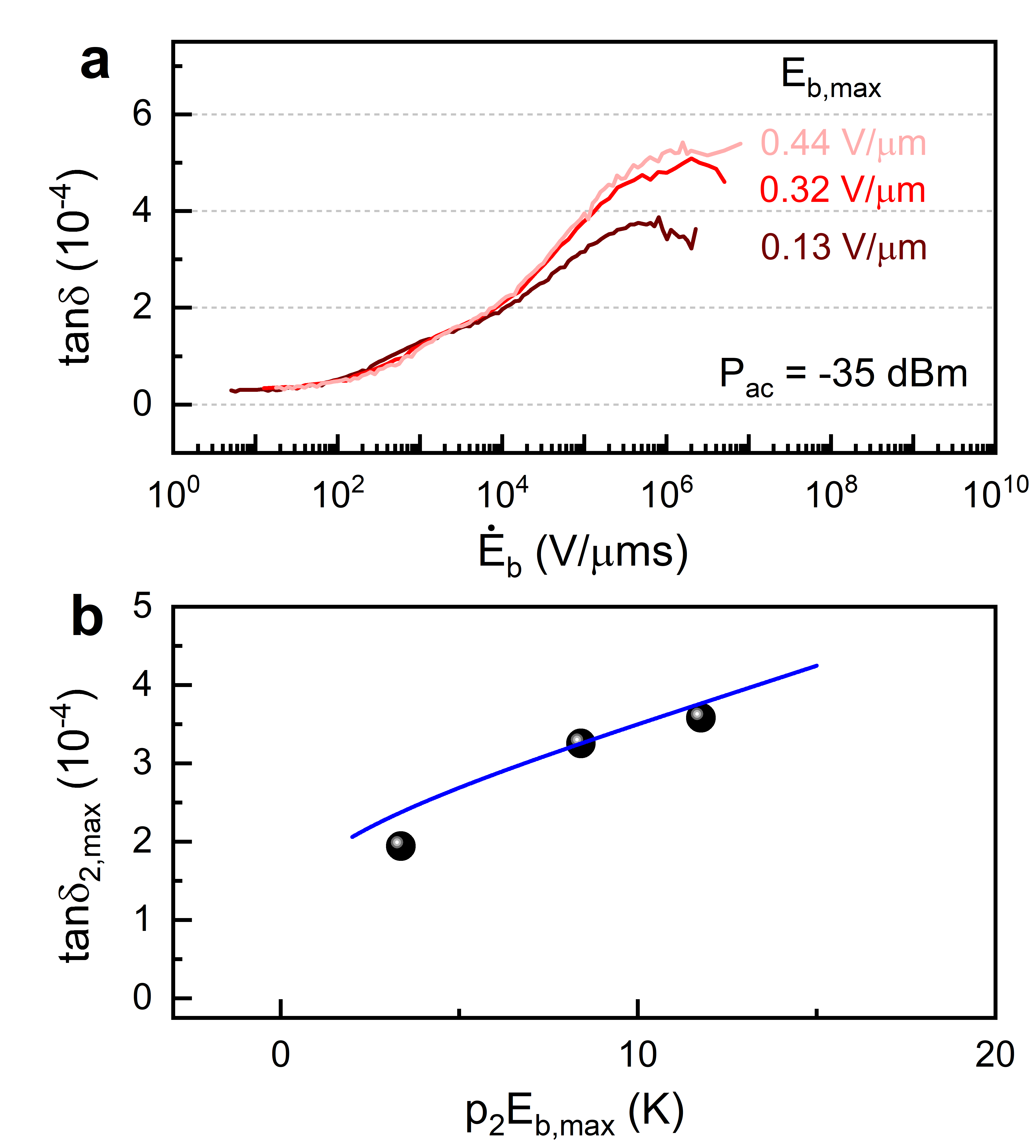}\vspace{-3mm}
  \caption{\textbf{Bias amplitude dependence of the maximum loss.}
\textbf{a}~$\tan\delta$ measured as a function of $\dot{E}_{\rm b}$ for different bias field amplitudes $E_{\rm b,max}$. \textbf{b}~The measured maximum excess loss $\tan\delta_{2,\rm max}$ as a function of $p_2 E_{\rm b,max}$, proportional to the equilibrium DOS of the second type of TLSs at this energy, $\rho^{}_{2,{\rm eq}}(p_2 E_{\rm b,max})$. The blue line shows a (weak) logarithmic dependence of the measured $\tan\delta_{2,\rm max}$, which indicates that we most likely measured in the regime where $p_2 E_{\rm b,max}$ is in the vicinity of or outside the gap edge. \vspace{-4mm}}
  \label{Figure:4}
\end{figure}

\noindent\textbf{Comparison to the two-TLS model.} From the calculation of $\tan\delta_2$ we can extract information that allows us to examine the predictions of the two-TLS model. First, the ratio $\Gamma^{(2)}_{1,\rm m}/\Gamma^{(1)}_{1,\rm m}$ between the relaxation rates of the two types of TLSs is equal to $(\gamma_2/\gamma_1)^2$, where $\gamma_1$ and $\gamma_2$ are the couplings of the two types of TLSs to phonons. Using the relaxation rates found above, we obtain $\gamma_2/\gamma_1 \approx 12$, which is slightly smaller than expected within the two TLS model\cite{Schechter2013,Gaita-Arino2011}, but nevertheless describes distinct coupling strengths to phonons differing by an order of magnitude. Second, the ratio between the equilibrium DOS of the two types of TLSs at the resonance energy $\hbar\omega_0\approx 0.25\,k_{\rm B}$K can be found from $A/\tan\delta_0=p^{2}_{2}\rho^{}_{2,{\rm eq}}(\hbar\omega_0)/(p^{2}_{1}\rho^{}_{1})\approx 10^{-4}$. Together with the dipole moments $p_1=11\,$D and $p_2=110\,$D, we find $\rho^{}_{2,{\rm eq}}(\hbar\omega_0)/\rho^{}_{1}\approx 10^{-6}$. This is consistent with the general success of the STM in describing the low-temperature universality in some acoustic and thermodynamic properties of amorphous solids, since phonon attenuation is dominated by the standard (weakly interacting) TLSs at energies where $\gamma^{2}_{2}\rho^{}_{2,{\rm eq}}(E)\ll\gamma^{2}_{1}\rho^{}_{1}$. Lastly, by estimating the largest energy of standard TLSs as $E_{\rm{max}}\approx 10\,k_{\rm B}$K, from the parameter $B$ one obtains $\rho^{}_{1}u\approx 4\times 10^{-2}$. According to the two-TLS model, the tunneling strength is given by $C_0 = \rho^{}_{1}u\cdot (\gamma_1/\gamma_2)$, which for $\gamma_1/\gamma_2\approx 1/12$ gives $C_0 \approx 3\times 10^{-3}$. This agrees with the universally small value of $C_0 \sim 10^{-4} - 10^{-3}$ which is experimentally found to hold across a wide range of different amorphous solids~\cite{Pohl2002}.

Further comparison with the two-TLS model is achieved by studying the bias rate dependence of the loss at various bias amplitudes. This is shown in Fig.~\ref{Figure:4}a for driving power of $-35\,$dBm. It reveals a clear bias amplitude dependence at high bias rates, indicating an energy-dependent DOS for the second TLS type. At the highest bias rates $\dot{E}_{\rm b}\gg De^{2C}$, where $De^{2C}=\Gamma_{1,\rm m}(E_{\rm{max}})E_{\rm{b,max}}\sim 5\times 10^{4}$V$/\mu$m$\,$s, the gap reconstruction in the non-equilibrium DOS $\rho_2(E,\dot{E}_{\rm b})$ is negligible. Thus $\rho_2(E,\dot{E}_{\rm b}\gg\Gamma_{1,\rm m}(E_{\rm{max}})E_{\rm{b,max}})$ approaches the equilibrium DOS $\rho^{}_{2,\mathrm{eq}}(E=\hbar\omega_0+p_2E_{\rm{b,max}})$ at the highest energy from which TLSs are brought into resonance. The excess loss at the highest bias rate is therefore $\tan\delta_{2,\rm{max}}\approx\pi p^{2}_{2}\rho_{2,{\rm eq}}(\hbar\omega_0+p_2E_{\rm b,max})/(3\varepsilon)$ and thus proportional to the equilibrium DOS at this energy. Within the two-TLS model, $\rho^{}_{2,{\rm eq}}(E)$ is predicted to have a power-law energy dependence below the maximum energy of the standard TLSs $E_{\rm{max}}$, followed by a logarithmic behavior at higher energies. The former results from interactions between the two types of TLSs, whereas the latter is a result of interactions among the strongly interacting TLSs~\cite{Burin1995,Churkin2014_2}. With a dipole moment $p_2=110\,$D, and bias amplitudes $E_{\rm b,max}=0.13\,$V/$\mu$m, $0.32\,$V/$\mu$m and $0.44\,$V/$\mu$m, the corresponding energies $\hbar\omega_0+p_2 E_{\rm b,max}\approx p_2 E_{\rm b,max}$ are estimated as $3\,k_{\rm B}$K, $8\,k_{\rm B}$K and $12\,k_{\rm B}$K. Figure~\ref{Figure:4}b shows the maximum excess loss $\tan{\delta}_{2,{\rm max}}$ as a function of $p_2 E_{\rm b,max}$, which is a manifestation of the equilibrium DOS of the second TLS type at this energy. Our measurements are most likely taken in the regime where $p_2 E_{\rm b,max}$ is the vicinity of or outside the gap edge, namely in the logarithmic dependence domain.
\\

\noindent\textbf{Discussion}\\ 

Using a fast-swept bias field in addition to microwave fields in a resonator, we have studied the non-equilibrium loss in amorphous silicon. The data suggests the existence of two types of TLSs. At low bias rates the dielectric loss is determined by standard TLSs, which are weakly coupled to phonons. At high bias rates the dielectric loss deviates from the STM predictions, but agree well with the two-TLS model used. Analysis of the bias rate dependent loss indicates a gap in the DOS and the fitting yields a large electric dipole moment for the second TLS type. Recent works~\cite{Carruzzo2020,Carruzzo2021} have examined the broad distribution of the coupling of the TLSs to the phonons due to polaron effect~\cite{Agarwal2013}, and the TLS-TLS interactions could lead to the reduction of TLS DOS. However, the bias application cannot release TLSs suppressed by the polaron effect because the polaron bath is formed by phonons almost instantaneously. Furthermore, a single continuously distributed TLS DOS cannot account for our experimental observations explained by two moments and relaxation parameters. Technically, the fittings of the data for the entire domain of driving powers and bias rates contain eight parameters (see Fig.~\ref{Figure:2}c). These include the dipole moments $p_1\approx 11\,$D, $p_2\approx 110\,$D and relaxation rates $\Gamma^{(1)}_{1,\rm m}\approx 5.7\,$MHz, $\Gamma^{(2)}_{1,\rm m}\approx 800\,$MHz for the two types of TLSs, and the four parameters $A$, $B$, $C$ and $D$ used to fit the unsaturated excess loss at low and intermediate powers. The accuracy of these parameters depends on the details of the LZ analysis and the fitting precision is better than 10\%. Also, our main results do not depend on the preciseness of these parameters, since the distinct dipole moments and relaxation rates of the two kinds of TLSs result from the distinct driving power dependence of the two contributions to the loss, below and above the intrinsic loss tangent $\tan\delta_0$. Experimental data consistent with the existence of two types of TLSs, characterized by distinct couplings to phonons, was previously reported~\cite{Matityahu2016,Kirsh2017}. Our results here go beyond previous experiments as data attests directly to the presence of two types of TLSs, the bimodality of their couplings to phonons, and their electric dipole interaction strengths.

The location of the second TLS type (corresponding to $p_2$) must be in the film, similar to the standard type. The parallel plate capacitor geometry allows for measuring loss from TLSs from the dielectric within the capacitor alone, effectively eliminating TLS contributions to the loss elsewhere because electric fields are negligible outside of the capacitor. The second TLS type, similar to the standard TLSs, is believed to be uniformly distributed throughout the film similar to the first type, because TLSs in the bulk film have a 100\% participation ratio, whereas those residing on the metal-dielectric interface in the capacitor would have a participation ratio on the order of 1\% (using a thickness of a few nanometers for the interface, as is standard and expected). 

The dipole moment of $110\,$D extracted for the second type of TLSs is anomalously large and it only appears out of equilibrium due to a gap in its density of states. One likely candidate for this TLS is tunneling nanoclusters of atoms, see e.g., Ref~\cite{Lubchenko2001}. Insights of the two types of TLSs gleaned from the experiment are of general interest to understand TLS properties in amorphous solids, and of increasing importance due to their impact in quantum information science. The fast bias technique has been previously used to study the non-equilibrium loss for a couple different materials: silicon nitride\cite{Khalil2014} and alumina\cite{Matityahu2019}. However, we have not yet seen excess loss in these materials. The two-TLS model is silent with respect to the electric dipole moments of the TLSs, which can vary between materials. A recent study suggests that mechanical and dielectric loss in amorphous silicon originates from two different types of TLSs~\cite{Molina2021}. Our measurement here is sensitive to the size of the electric dipole moment of the TLSs, which is plausibly material dependent. To check the generality of the second TLS type in other amorphous solids, one possibility is to repeat our non-equilibrium protocol, but measure both the dielectric and acoustic responses, as large acoustic response out of equilibrium is expected irrespective of the value of the electric dipole moment. Alternatively, one could also measure the thermal conductivity, which is proportional to square of the TLS-phonon coupling, and compare the results with and without a rapidly varying bias field. Such studies would allow detailed characterization of TLSs at low and high energies. This could clearly establish if high-energy TLSs will appear from high sweep rates in qubits and reduce their coherence, as well as help to uncover an appropriate microscopic model.
\\

\noindent\textbf{Methods}\\
\noindent\textbf{Experimental setup}\\
The resonators are fabricated with Al/a-Si/Al trilayer films on highly resistive silicon substrates. The bottom aluminum layer ($100\,$nm) and the top aluminum layer ($250\,$nm) are sputtered via DC sputtering. The low-stress amorphous silicon layer ($250\,$nm) is grown by PECVD at $100^{\circ}$C. The film can have oxygen- and hydrogen-based impurities related to the low growth temperature and $SiH_4$ used in the growth, either of which might be responsible for the moments in $p_1$ (see discussion above). The resonator (Fig.~\ref{Figure:1}a) is defined by photolithography and subsequent etching. It consists of four equal bridge parallel-plate capacitors (Figs.~\ref{Figure:1}a and ~\ref{Figure:1}b), which is modified from that in ref.31. Assuming the dielectric constant for the amorphous silicon is 11.5, the capacitor has a total capacitance $C = 1.2\,$pF ($C1-C4 = C$). The vias are etched by $SF_6$ to connect the capacitor to the meander inductor. The resonator is inductively coupled to the coplanar transmission line for standard transmission measurements of $S_{21}$. The measurements were performed at the base temperature of the dilution refrigerator ($T_{\rm base} \approx 20\,$mK). A dc bias field, $E_{\rm b}=V_{\rm b}/2d$, where $d=250\,$nm is the thickness of the amorphous silicon dielectric, is applied across the capacitors. The bridge layout is effective to isolate the microwave resonance energy from the bias field input port. A triangular bias voltage is applied. The time dependence of the loss for a resonator when a periodic bias field is applied is studied in Ref.\cite{Khalil2014}. The maximum bias is $E_{\rm b,max} = 0.44\,$V/$\mu$m, and the fastest bias frequency is $f_b = 4.5\,$MHz. The resonance frequency $f_0 = 5.1\,$GHz. $S_{21}$’s are measured as a function of photon number, $n$ and bias rate, $\dot{E}_{\rm b}$. The corresponding loss $\tan{\delta}$, equal to the inverse resonator quality $1/Q_{\rm i}$, is found from each $S_{21}$ measurement. See Supplementary Note 1 for more wiring and attenuation details in the measurement setup.\\

\noindent\textbf{Loss due to LZ transitions within the STM}\\
The Hamiltonian of a single TLS driven by the resonator electric field $\vec{E}_{\rm res}(t)=\vec{E}_{\rm ac}\cos(\omega_0 t)$ and by a time-dependent bias electric field $\vec{E}_{\rm b}(t)$ is $H=(1/2)(\Delta(t)\sigma_z+\Delta_0\sigma_x)-\vec{p}\cdot\vec{E}_{\rm ac}\cos(\omega t)\sigma_z$, where $\sigma_i$ ($i=x$, $y$, $z$) are the Pauli matrices, $\vec{p}$ is the TLS dipole moment, $\Delta(t)=\Delta(0)-2\vec{p}\cdot\vec{E}_{\rm b}(t)$ is the asymmetry energy tuned by the bias field and $\Delta_0$ is the tunneling energy. In the instantaneous eigenbasis of the TLS, the corresponding Hamiltonian is 
\begin{equation}
\label{Eq:2}
H=\frac{1}{2}E(t)\sigma_z-\vec{p}\cdot\vec{E}_{\rm ac}\cos(\omega_0 t)\left(\frac{\Delta(t)}{E(t)}\sigma_z-\frac{\Delta_0}{E(t)}\sigma_x\right),
\end{equation}
where $E(t)=\sqrt{\Delta^{2}(t)+\Delta^{2}_{0}}$ is the TLS energy splitting, assumed to be slowly varying on the time scale $2\pi/\omega_0$. A TLS with $\Delta^{}_{0}<\hbar\omega_0$ can be swept through resonance at time $t^{}_{0}$ for which $E(t^{}_{0})=\hbar\omega_0$. Near this resonance the energy splitting can be expanded as $E(t)\approx\hbar\omega_0+\hbar v(t-t^{}_{0})$~\cite{Khalil2014,Burin2013}, where $v=\dot{E}(t^{}_{0})/\hbar=v^{}_{0}\sqrt{1-\left(\Delta^{}_{0}/\hbar\omega_0\right)^{2}}\cos\theta$ with $\theta$ the angle between $\vec{p}$ and $\vec{E}_{\rm b}$, and $v^{}_{0}=(2p/\hbar)\dot{E}_{\rm{b}}(t^{}_{0})$ the maximum bias rate. 

The TLS absorption in the absence of the bias field can be separated into the so-called relaxation absorption, arising from the longitudinal term ($\propto\sigma_z$) in the brackets of Eq.~(\ref{Eq:2}), and the resonant absorption resulting from the transverse coupling ($\propto\sigma_x$)~\cite{Phillips1987,Hunklinger1986,Jackle1976,Carruzzo1994}. In the regime $\omega_0\gg k_{\rm B}T/\hbar\gg\Gamma_1, \Gamma_2$ considered in this paper ($\Gamma_1$ and $\Gamma_2$ are the TLS relaxation and decoherence rates, respectively), the resonant mechanism dominates and the longitudinal term in Eq.~(\ref{Eq:2}) can be neglected. The TLS dipole moment induced by the resonator field can  therefore be written as $\langle p(t)\rangle=-p\cos\theta(\Delta_0/\hbar\omega_0)\langle\sigma_x(t)\rangle=\Re\left[\chi E_{\rm ac}e^{-i\omega_0 t}\right]$, where $\chi=\chi'+i\chi''$ is the TLS electric susceptibility. The imaginary part of the susceptibility, $\chi''$, yields the imaginary part $\varepsilon''$ of the dielectric constant due to TLSs upon averaging over the ensemble of TLSs ($\chi'$ gives the shift in the real part of the dielectric constant due to TLSs, which is small compared to the material dielectric constant $\varepsilon$). The loss tangent $\tan\delta=\varepsilon''/\varepsilon$ is thus proportional to the out of phase ($\propto\sin(\omega_0 t)$) component of $\braket{\sigma_x(t)}$. To calculate this component, we transform to the frame of reference rotating around the $z$ axis with frequency $\omega_0$, using the unitary transformation $U_{\rm R}=e^{i\omega_0 t\sigma_z/2}$. Since $U_{\rm R}\sigma_x U^{\dag}_{\rm R}=\cos(\omega_0 t)\sigma_x-\sin(\omega_0 t)\sigma_y$, in the rotating frame of reference the relation $\chi''=(\hbar\Omega_{\rm R}/E^{2}_{\rm ac})\braket{\sigma_y}$ holds~\cite{Carruzzo1994}, where $\Omega_{\rm R}=\Omega_{\rm R0}\cos\theta(\Delta_0/\hbar\omega_0)$ is the TLS Rabi frequency, with its maximum value $\Omega_{\rm R0}=pE_{\rm{ac}}/\hbar$. Moreover, application of the rotating wave approximation yields the LZ Hamiltonian 
\begin{equation}
\label{Eq:3}
H_{\rm LZ}=\frac{\hbar}{2}\left[v(t-t_0)\sigma_z+\Omega_{\rm R}\sigma_x\right],
\end{equation}
governing the dynamics of a TLS resonant passage~\cite{Burin2013}. The probability for photon absorption in a transition is the famous LZ probability for an adiabatic transition from the initial state $\ket{g,n}$ to the final state $\ket{e,n-1}$ [Fig.~\ref{Figure:1}(d)], $P_{\rm ad}=1-e^{-\pi\Omega^{2}_{\rm R}/(2v)}$. Finite TLS relaxation and decoherence rates $\Gamma_1$ and $\Gamma_2$ can be taken into account by means of the Bloch equations within the rotating frame of reference~\cite{Carruzzo1994}. The Hamiltonian~(\ref{Eq:3}) can be written as the Hamiltonian of a spin-1/2 particle in a magnetic field, $H_{\rm LZ}=-\vec{S}\cdot\vec{B}$, where $\vec{S}=\hbar\vec{\sigma}/2$ and $\vec{B}=-(\Omega_{\rm R},0,v(t-t_0)/\hbar)$. The corresponding Bloch equations read $d\langle\vec{S}\rangle/dt=\langle\vec{S}\rangle\times\vec{B}-(\Gamma_2\langle S_x\rangle,\Gamma_2\langle S_y\rangle,\Gamma_1(\langle S_z\rangle-S_{z0}))$, where $S_{z0}=(\hbar/2)\tanh(\hbar\omega_0/2k_{\rm B}T)$. In our calculations we neglect pure dephasing of the resonant TLSs, such that $\Gamma_2=\Gamma_1/2$. Assuming TLS relaxation into the phonon bath, one has $\Gamma_1(\Delta_0)=\Gamma_{1,\rm m}(\Delta_0/\hbar\omega_0)^2$, where $\Gamma_{1,\rm m}$ is the maximum relaxation rate for resonant TLSs with energy splitting $E=\hbar\omega_0$. If the bias duration is longer than the TLS relaxation time $T_1=1/\Gamma_1$, the steady state solution depends on time via the detuning from resonance $\delta E=v(t-t_0)$, and $\varepsilon''$ can be obtained by averaging $\chi''$ over the ensemble of TLSs with the distribution function $P(E,\Delta_0)=\rho E/(\Delta_0\sqrt{E^{2}-\Delta^{2}_{0}})$, where $\rho$ is the TLS DOS, assumed to be energy-independent within the STM. The integration over energies can then be replaced by integration over time, leading to the expression
\begin{equation}
\label{Eq:4}
\frac{\tan\delta}{\tan\delta_0}=3\int^{1}_{0}dyy^{2}\int^{1}_{0}\frac{xdx}{\sqrt{1-x^2}}\frac{2v}{\pi\Omega_{\rm R}}\int^{\infty}_{-\infty}\braket{\sigma_y(t)}dt,
\end{equation}
where $\tan\delta_0=\pi\rho p^{2}/(3\varepsilon)$ is the intrinsic loss tangent and the integration variables are $x=\Delta_0/(\hbar\omega_0)$ and $y=\cos\theta$ (in terms of these variables $v=v_0\sqrt{1-x^2}\,y$ and $\Omega_{\rm R}=\Omega_{\rm R0}xy$). This loss is a function of the parameters $\xi=2v_0/(\pi\Omega^{2}_{\rm{R}0})$ and $\Gamma_{1,\rm m}/\Omega_{\rm{R}0}$, and can be calculated by a numerical integration of the Bloch equations.

In the regime $\Gamma_{1,\rm m}/\Omega_{\rm R0}\ll 1$, dissipation can be neglected for $\xi\gg\Gamma_{1,\rm m}/\Omega_{\rm R0}$, and one obtains the expression for the dielectric loss tangent~\cite{Burin2013}
\begin{equation}
\label{Eq:5}
\frac{\tan\delta}{\tan\delta_0}=3\int^{1}_{0}dyy^{2}\int^{1}_{0}\frac{xdx}{\sqrt{1-x^2}}\frac{2v}{\pi\Omega^{2}_{\rm R}}\left[1-e^{-\pi\Omega^{2}_{\rm R}/(2v)}\right].
\end{equation}
By performing numerical integration over $x$ and $y$, one obtains the normalized loss $\tan\delta/\tan\delta_0$ as a function of the dimensionless bias rate $\xi$, approaching unity at $\xi\gg 1$, as shown in Fig.~\ref{Figure:1}(d).\\

\noindent\textbf{Loss due to the second type of TLSs}\\
Let us derive the expression for the non-equilibrium DOS of the second type of TLSs, $\rho^{}_{2}(E,\dot{E}_{\rm b})$, under the assumption that their energy-dependent equilibrium DOS results from interactions between the two types of TLSs, as motivated by the Efros-Shklovskii coulomb gap~\cite{Efros1975,Baranovski79} and discussed in Refs.~\cite{Schechter2013,Churkin2014_2}. In equilibrium, the DOS is reduced by interaction with standard TLSs that break the Efros-Shklovskii stability criterion $E+E'-2u/R^{3}>0$, where $E'$ is the energy of the standard TLS, $u/R^3$ is the mutual TLS-TLS interaction and $R$ is the distance between the TLSs. We note that interactions exist also among the second type of TLSs, but since they are scarce at low energies, the dominating interactions are with the standard TLSs. As discussed in Refs.~\cite{Schechter2013,Churkin2014_2}, at higher energies interactions among the second type of TLSs lead to weaker logarithmic energy dependence of the DOS [see Fig.~\ref{Figure:4}]; here we neglect these interactions and consider the gap in the DOS at low energies due to interactions between the two types of TLSs. In the non-equilibrium situation of our experiment, only those standard TLSs that are capable of changing their state during the sweep time $1/(2f_{\rm b})$ will contribute to the reconstruction of the gap in the DOS. For $f_{\rm b}\ll\Gamma_1$, where $\Gamma_1$ is a typical relaxation rate of standard TLSs, the DOS $\rho^{}_{2}(E,f_{\rm b}\ll\Gamma_1)$ should approach the DOS at equilibrium, $\rho^{}_{2,\rm eq}(E)$, discussed in Ref.~\cite{Schechter2013}. On the other hand, in the limit of instantaneous sweep, $f_{\rm b}\gg\Gamma_1$, the non-equilibrium DOS $\rho^{}_{2}(E,f_{\rm b})$ should satisfy the condition $\rho^{}_{2}(E,f_{\rm b}\gg\Gamma_1)\approx\rho^{}_{2,\rm eq}(E+p_{2}E_{\rm b,max})$. Taking into account the fraction $1-e^{-\Gamma_1/(2f_{\rm b})}$ of standard TLSs that can flip during time $1/(2f_{\rm b})$, where $\Gamma_1=\Gamma_1(E',\Delta_0)$ is the relaxation time of standard TLSs with energy $E'$ and tunneling energy $\Delta_0$, we obtain the expression
\begin{widetext}
\begin{align}
\label{Eq:6}\rho^{}_{2}(E,f_{\rm b})&\approx\prod^{}_{d^{3}R}\Bigg[1-d^{3}R\int^{E_0}_{0}\rho^{}_{1}dE'\int^{E'}_{\Delta^{}_{0,\mathrm{min}}}\frac{d\Delta^{}_{0}}{\Delta^{}_{0}}\Theta\left(\frac{2u}{R^{3}}-E-E'\right)\left(1-e^{-\frac{\Gamma_1}{2f_{\rm b}}}\right)\Bigg]\rho^{}_{2,\mathrm{eq}}(E+p_{2}E_{\mathrm{b,max}})\nonumber\\
&\approx\mathrm{exp}\left[-\rho_{1}\int^{E_0}_{0}dE'\int^{E'}_{\Delta^{}_{0,\mathrm{min}}}\frac{d\Delta^{}_{0}}{\Delta^{}_{0}}\int d^{3}R\,\Theta\left(\frac{2u}{R^{3}}-E-E'\right)\left(1-e^{-\frac{\Gamma_1}{2f_{\rm b}}}\right)\right]\rho^{}_{2,\mathrm{eq}}(E+p_{2}E_{\mathrm{b,max}})\nonumber\\
&=\mathrm{exp}\Bigg[\rho^{}_{1}\int^{E_0}_{0}dE'\int^{E'}_{\Delta^{}_{0,\rm{min}}}\frac{d\Delta^{}_{0}}{\Delta^{}_{0}}\int d^{3}R\,\Theta\left(\frac{2u}{R^{3}}-E-E'\right)e^{-\frac{\Gamma_1}{2f_{\rm b}}}\Bigg]\rho^{}_{2,\mathrm{eq}}(E),
\end{align} 
\end{widetext}
where $\Theta(x)$ is the step function and $E_0=\min\{E_{\rm{max}},\hbar\omega_0+p_{2}E_{\rm b,max}\}$ with $E_{\rm{max}}\sim 10\,$K being the maximum energy of standard TLSs. This expression fulfills both limiting conditions discussed above. The exponent in the last line of Eq.~(\ref{Eq:6}) gives the enhancement of the equilibrium DOS $\rho^{}_{2,\rm eq}(E)$ due to standard TLSs with relaxation rates $\Gamma_1\lesssim f_{\rm b}$. Using $\dot{E}_{\rm b}=2f_{\rm b}E_{\rm b,max}$ and assuming relaxation of standard TLSs to be dominated by interaction with phonons, corresponding to the rate $\Gamma_{1}(E',\Delta_0)=\Gamma_{1,\rm m}(E')(\Delta_0/E')^2$ with $\Gamma_{1,\rm m}(E')\propto E'^{3}$~\cite{Phillips1987}, we obtain
\begin{widetext}
\begin{align}
\label{Eq:7}\rho^{}_{2}(E,\dot{E}_{\mathrm b})&=\rho^{}_{2,\mathrm{eq}}(E)\,\mathrm{exp}\Bigg[\frac{8\pi}{3}\rho^{}_{1}u\int^{E_0}_{0}\frac{dE'}{E+E'}\int^{\ln(E'/\Delta^{}_{0,\mathrm{min}})}_{0}dx\,e^{-\Gamma_{1,\rm m}(E')(\Delta_{0,\mathrm{min}}/E')^2(E_{\mathrm{b,max}}/\dot{E}_{\mathrm b})e^{2x}}\Bigg],
\end{align}
\end{widetext}
where $x=\ln(\Delta_0/\Delta^{}_{0,\rm{min}})$. Since the major contribution to the integral over the tunneling amplitudes $\Delta_0$ comes from slow TLSs with small $\Delta_0$, we can further simplify the last expression by replacing $E'$ by $E_{\rm max}$ in the last integral and in the upper limit. This results in the expression
\begin{widetext}
\begin{align}
\label{eq:DOS4}\rho^{}_{2}(E,\dot{E}_{\mathrm b})&\approx\rho^{}_{2,\mathrm{eq}}(E)\,\mathrm{exp}\Bigg[\frac{8\pi}{3}\rho^{}_{1}u\ln\left(1+\frac{E_0}{E}\right)\int^{\ln\left(\frac{E_{\rm max}}{\Delta^{}_{0,\rm min}}\right)}_{0}dx\,e^{-\Gamma_{1,\rm max}(E_{\rm max})\left(\frac{\Delta_{0,\rm min}}{E_{\rm max}}\right)^2\frac{E_{\mathrm{b,max}}}{\dot{E}_{\mathrm b}}e^{2x}}\Bigg].
\end{align}
\end{widetext}
At low power driving, where the second type of TLSs is non-saturated, the excess loss is $\tan\delta_{2,0}(\dot{E}_{\rm b})=\pi p^{2}_{2}\rho^{}_{2}(\hbar\omega_0,\dot{E}_{\rm b})/(3\varepsilon)$ and thus takes the form of Eq.~(\ref{Eq:1}). This loss serves as the intrinsic loss for the second TLS type, and is used for scaling the excess loss in Fig.~(\ref{Figure:3}). The LZ analysis described above is then carried out for the normalized excess loss $\tan\delta_2/\tan\delta_{2,0}$ at all driving powers.

\section{Data availability}
The data in the main text and Supplementary Materials are available from the corresponding authors upon request.

\section{Acknowledgments}
L.Y. would like to thank F. C. Wellstood and S. Dutta for their discussions and support. S.M. acknowledges support by the A. von Humboldt foundation. Y.J.R. acknowledges Lawrence Livermore National Laboratory operated by Lawrence Livermore National Security, LLC, for the U.S. Department of Energy, National Nuclear Security Administration under Contract DE-AC52-07NA2. A.L.B. and A.M. acknowledge the support by Carrol Lavin Bernick Foundation Research Grant (2020-2021), NSF CHE-1900568 grant and LINK Program of the NSF and Louisiana Board of Regents. M.S. acknowledges support from the Israel Science Foundation (Grant No. 2300/19).

\section{Author Contributions}
L.Y., Y.J.R., C.H. and K.D.O. designed and performed the experiments, L.Y., K.D.O. and S.M. analysed the data and wrote the paper. L.Y. and Y.J.R. processed the samples.  S.M., M.S. A. M. and A.L.B. developed and carried out the theoretical work and numerical modeling. All authors discussed the results and commented on the manuscript.

\section{Competing interests}
The authors declare no competing interests.

\end{document}



\title{SUPPLEMENTARY MATERIAL \\ Experimentally revealing anomalously large dipoles in a quantum-circuit dielectric}


\author{Liuqi Yu}
\affiliation{Laboratory for Physical Sciences, University of Maryland, College Park, MD 20740, USA}
\affiliation{Department of Physics, University of Maryland, College Park, MD 20742, USA}

\author{Shlomi Matityahu}
\affiliation{Department of Physics, Ben-Gurion University of the Negev, Beer Sheva 84105, Israel}
\affiliation{Institut f\"ur Theorie der Kondensierten Materie, Karlsruhe Institute of Technology, 76131 Karlsruhe, Germany}

\author{Yaniv J. Rosen}
\affiliation{Lawrence Livermore National Laboratory, Livermore, California 94550 USA}

\author{Chih-Chiao Hung}
\affiliation{Laboratory for Physical Sciences, University of Maryland, College Park, MD 20740, USA}
\affiliation{Department of Physics, University of Maryland, College Park, MD 20742, USA}

\author{Andrii Maksymov}
\affiliation{Department of Chemistry, Tulane University, New Orleans, LA 70118, USA}

\author{Alexander L. Burin}
\affiliation{Department of Chemistry, Tulane University, New Orleans, LA 70118, USA}

\author{Moshe Schechter}
\affiliation{Department of Physics, Ben-Gurion University of the Negev, Beer Sheva 84105, Israel}

\author{Kevin D. Osborn}
\affiliation{Laboratory for Physical Sciences, University of Maryland, College Park, MD 20740, USA}
\affiliation{Joint Quantum Institute, University of Maryland, College Park, MD 20742, USA}

\textbf{\large{Experimentally revealing anomalously large dipoles in a quantum-circuit dielectric, \\Yu \textit{et al.}}}
\thispagestyle{empty}
\clearpage
\let\newpage\relax

\maketitle

\tableofcontents
\setcounter{page}{\pagenumbaa}
\thispagestyle{plain}
\clearpage

\section{Additional details of the experimental setup} \label{sec:Setup}

\begin{figure}[htb!]
\centering
\includegraphics[width=0.7\columnwidth]{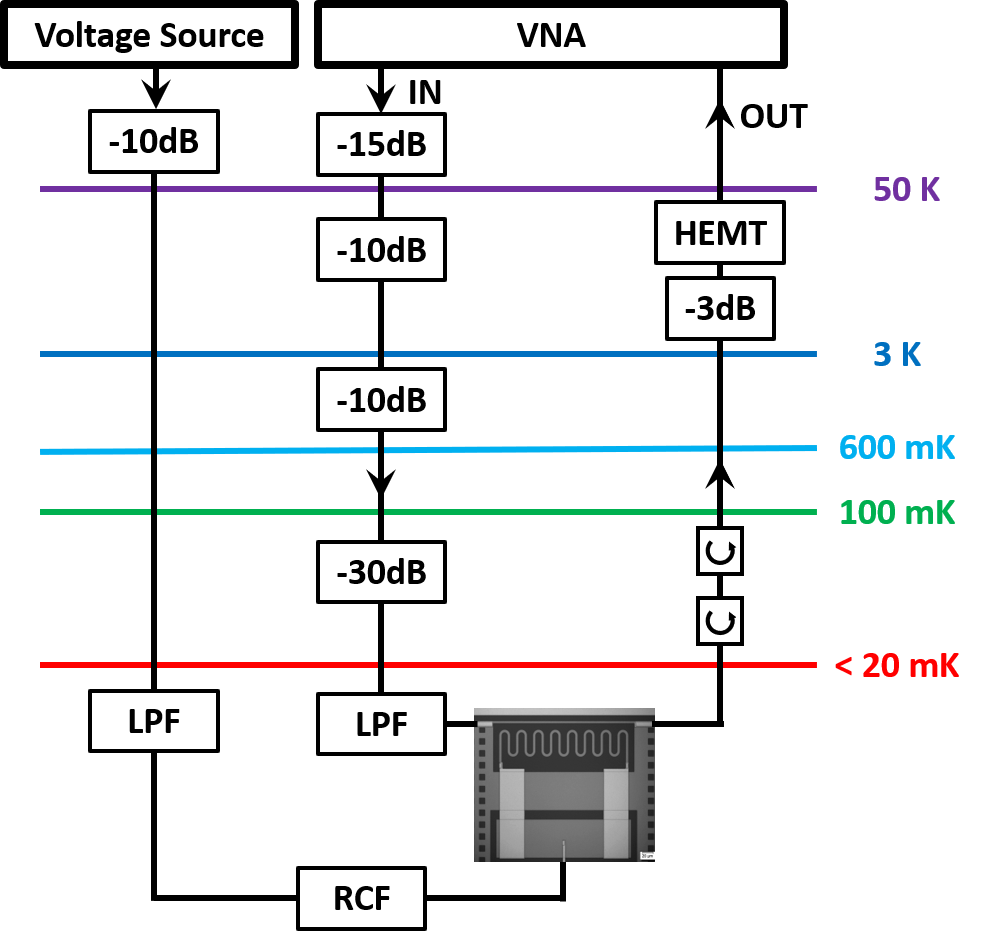}
\caption{
\textbf{Schematic of the experimental wiring for measuring the resonator.} }
\label{FigureSUP:1_1}
\end{figure}

The measurements are performed in a dilution refrigerator at a temperature of less than $20\,$mK. Supplementary Figure~\ref{FigureSUP:1_1} shows the schematic of the detailed wirings of the measurement setup. A vector network analyzer (VNA) is used to carry out the $S_{21}$ measurements. The input line is heavily filtered with multiple attenuators at different stages. An additional $12\,$GHz K\&L low-pass filter is added. The output signal is routed through two low-noise circulators ($4-12\,$GHz) which are thermally anchored at mixing chamber plate. The output signal is amplified by a high electron mobility transistor (HEMT) at $4\,$K. For the bias line, a $-10\,$dB attenuator is added at room temperature. A $12\,$GHz K\&L low-pass filter and a RC filter are placed at mixing chamber stage, which gives a cut-off frequency of $10\,$MHz. A time-varying bias voltage as a triangular waveform is applied. The maximum bias is $E_{\rm b} = 0.44\,$V$/\mu$ m, and the fastest applied bias frequency is $f_b = 4.5\,$ MHz. In previous related measurements, the time-dependent loss, particularly when the bias rate changes sign, was studied\cite{Khalil2014}. It is known that the change to the measured average loss is negligible. The base temperature of the dilution refrigerator remains below $20\,$mK throughout all measurements, such that no significant heating from bias leakage current is observed.  

\clearpage
\section{Input microwave power calibration}
To determine the input power on chip due to attenuation in the input line, throughput $S_{21}$’s (bypassing the mounted device) at various source powers have been measured at room temperature. This is particularly important to accurately determine the average photon number in the resonator. The attenuation depends on the signal frequency and the source power from the VNA. Four relatively high source power ($10\,$dBm, $0\,$dBm, $-10\,$dBm and $-20\,$dBm) are used to obtain sufficient signal to noise ratio. As shown in Supplementary Fig.~\ref{FigureSUP:1_2}a, linear fits are conducted on all four semilog $S_{21}$ plots. The frequency dependence is obtained by averaging the four slopes at different applied source powers. As shown in Supplementary Fig.~\ref{FigureSUP:1_2}b, the source power dependence of the input power at zero frequency is determined by plotting the intercepts as a function of the source powers. The obtained slope is one, which suggests any nonlinearities in the system associated with the power is negligible. Therefore, the eventual input power at mixing chamber can be expressed as 

\begin{figure}[htb!]
\centering
\includegraphics[width=1\columnwidth]{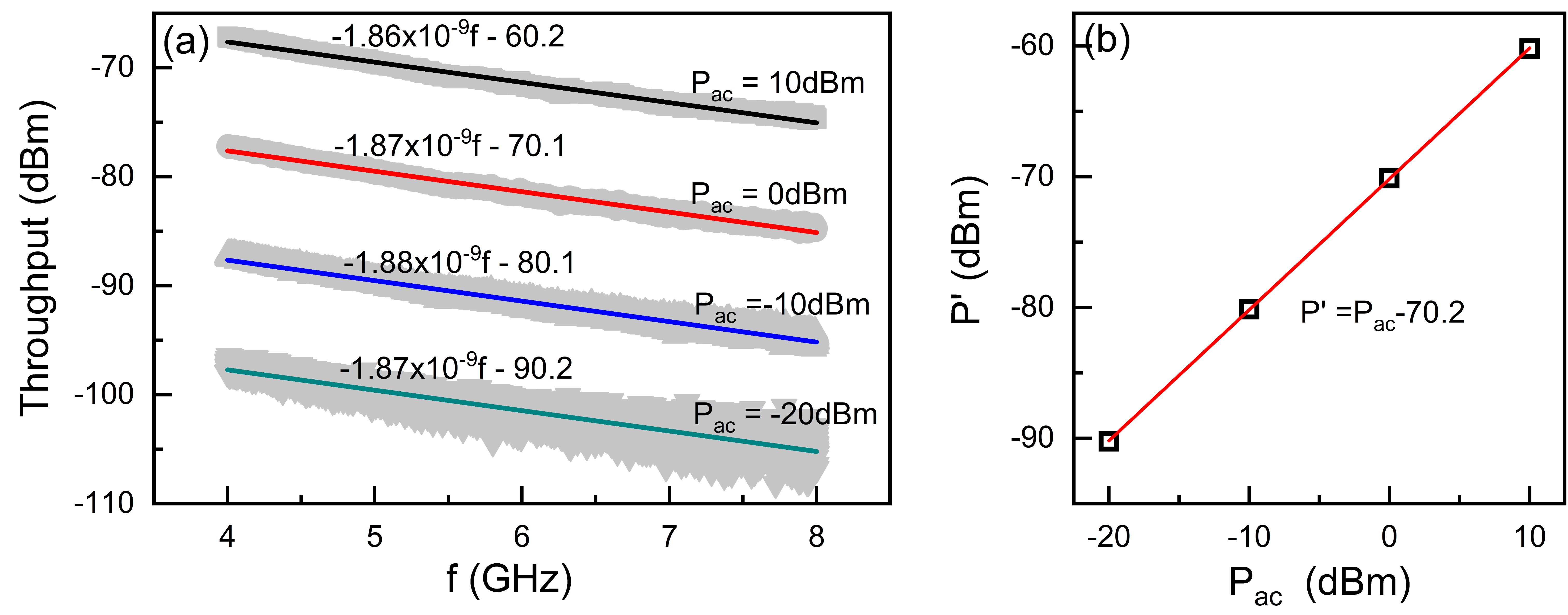}
\caption{
\textbf{a} Throughput calibration measurements of microwave source power at room temperature. \textbf{b} Plot and linear fit of the intercepts from the fits in \textbf{a}.}
\label{FigureSUP:1_2}
\end{figure}

\be
P = -1.87 \times 10^{-9}f + P_{\mathrm ac} - 70.2
\ee
	
This expression is used to calibrate the photon number on the measured resonator throughout the main manuscript. 

\clearpage

\clearpage
\section{Extraction of loss from $S_{21}$}

\begin{figure}[htb!]
\centering
\includegraphics[width=1\columnwidth]{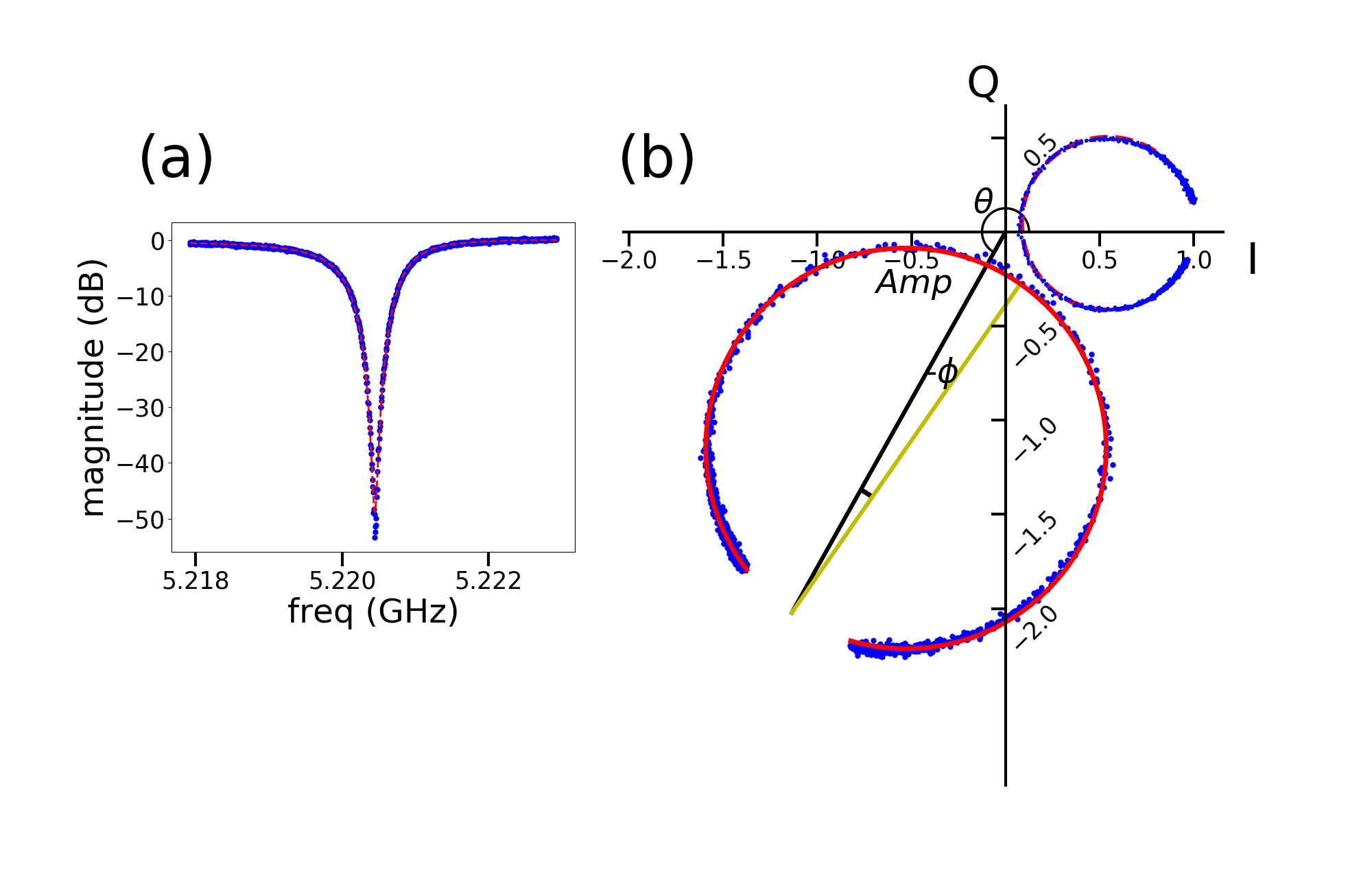}
\caption{
\textbf{a} An example of magnitude of $|S_{21}|$ vs frequency measured at $P_{\rm ac} = -29\,$dBm. \textbf{b} Raw $S_{21}$ data (blue dots) are plotted in IQ plane in the lower left. The red curve is a fitting of Supplementary Eq.~\ref{eq:S21}. The black line represents the scaling of amplitude and the rotation from $(1,0)$. The yellow line represents a $\phi$ rotation from a small impedance mismatch. We plot the data with adjusted amplitude and electric delays on the right.}
\label{FigureSUP:1_3}
\end{figure}

The quality factor extraction is following the fitting function\cite{Khalil2012} 

\be\label{eq:S21}
S_{21} = Amp \times e^{i\theta} \times \frac{1 - Q |\hat{Q}_{\mathrm e}^{-1}| e^{i\phi}}{1 + 2iQ\frac{\omega - \omega_0}{\omega_0}}.
\ee

The $Amp$ and $\theta$ are the adjusted parameters for the raw data due to the attenuations, amplifications, and electric delays in both input and output line, and $Q$ is the total quality factor. 
Because of the impedance mismatch at two ports, $\hat{Q}_{\rm e}$ is the external quality factor and also a complex number, and $\phi$ is the angle of $\hat{Q}_{\rm e}^{-1}$.
Fig.~\ref{FigureSUP:1_3}a shows an exemplary $S_{21}$ data (magnitude vs frequency) measured at $P_{\rm ac} = -29\,$dBm. The corresponding IQ plot of the data are shown in Fig.~\ref{FigureSUP:1_3}b: The lower left curve (blue dots) and the right curve (blue dots) are data before and after the adjusted $Amp$ and $\theta$.
From the fitting, we extract $Q_{\rm e} = 1/Re\{\hat{Q}_{\rm e}^{-1}\} \approx 7300$,  $Q_{\rm i} \approx 9000$, and $\phi = 5.7^{\circ}$ when the input photon number $n = 1422$.

\clearpage
\centerline{ \textbf{\large  Supplementary References } } 

\bigbreak